\documentclass[
    ,final            % use final for the camera ready runs
  ]
  {aipproc}

\layoutstyle{6x9}

\begin{document}

\title{MHD Simulations of Core Collapse Supernovae with Cosmos++}


\author{Shizuka Akiyama}{
  address={shizuka@slac.stanford.edu}
}

\author{Jay Salmonson}{
  address={salmonson1@llnl.gov}
}

\begin{abstract}

We performed 2D, axisymmetric, MHD simulations with Cosmos++ in order to
examine the growth of the magnetorotational instability (MRI) in core--collapse supernovae.
We have initialized a non--rotating 15 M$_\odot$ progenitor, 
infused with differential rotation and poloidal magnetic fields.  The collapse of the
iron core is simulated with the Shen EOS, and the parametric Ye and entropy
evolution.  The wavelength of the unstable mode in the post--collapse environment is
expected to be only $\sim$ 200 m.  In order to achieve the fine spatial resolution requirement, 
we employed remapping technique after the iron core has collapsed and bounced.

The MRI unstable region appears near the equator and angular momentum and entropy are
transported outward.  Higher resolution remap run display more vigorous overturns and stronger 
transport of angular momentum and entropy.  Our results are in agreement with the earlier
work by \citet{aki03} and \citet{ober09}.

\end{abstract}

\maketitle

%%%%%%%%%%%%%%%%%%%%%%%%%%%%%%%%%%%%%%%%%%%%
%% MAINMATTER
%%%%%%%%%%%%%%%%%%%%%%%%%%%%%%%%%%%%%%%%%%%%

\section{Introduction}

%Today with modern telescopes, we discover new supernovae almost at
%a rate of one per day.  Some of them are that of Type Ia supernovae, thermonuclear
%explosions of white dwarfs in binary systems, and others are that
%of core--collapse supernovae, explosions initiated by gravitational collapse
%of iron cores inside of massive stars.
%When a iron core collapses and a proto neutron star is formed, huge 
%amount of gravitational binding energy
%is liberated mostly in the form of neutrinos, mostly streaming away from the
%explosion site.  What happens between the formation of a proto neutron star
%and successful explosion is not well understood.  Currently multiple mechanisms are
%proposed, i.e. neutrino heating with background hydrodynamical instabilities, 
%acoustic powered, and MHD powered explosions.  It is, however, still immature 
%to draw a conclusion, and the topic continues to be central issue of the
%core--collapse supernovae explosion mechanism.

Neutron stars are spinning and magnetized, and magnetars are the extreme
neutron stars with magnetic fields > 10$^{14}$ G.  The GRBs associated with
ultra--relativistic outflow are considered to be generated by the MHD jets from the
rotating, magnetized engine; the so-called "Collapsar" or magnetar.  The close association of
those objects to core-collapse supernovae leads to the
question whether supernova explosions, too, are involved with rotation and magnetic fields.
If it is the case for most core collapse supernovae, the
modest initial magnetic fields have to be amplified more efficiently than linear wrapping
of the field lines or compression.  \citet{aki03} pointed out that the core collapse
environment is naturally unstable to the magnetorotational instability (MRI) \cite{BH91}, which
amplifies magnetic fields exponentially in linear regime.  Their 1D
study indicated that a small seed field in
the iron core is amplified to $\sim 10^{15-17}$ G within tens of milliseconds after bounce.
This prediction
has to be verified in numerical simulations because of the highly non--linear nature
of the problem.

\section{Numerical Models and Method}

The collapse of iron core is simulated by 2D, axisymmetric, MHD simulations
using Cosmos++ code.  Cosmos++ is a multidimensional, massively 
parallel,  Newtonian and fully general relativistic radiation-magneto-hydrodynamical code 
\cite{anni05}.  
We begin with a  non--rotating 15 M$_\odot$ progenitor model, 
s15s2b7 \cite{ww95}.  For rotation, we employ constant angular 
momentum profile:
$\Omega(r) = \Omega_{0} \frac{R^{2}}{r^{2} + R^{2}}$,
where $\Omega_{0}$ is the initial central angular velocity, and $R$ 
determines the degree of differential rotation.  Magnetic field
is obtained by taking curl of the magnetic potential
due to a current loop with a radius $D$ \cite{jackson75}.  The peak of
the magnetic field is set to a magnetic field strength $B_{0}$.  
We present a model with $\Omega_0 = 2.0$ [rad/s],
$R = 850$ km, $D = 400$ km, and $B_{0} = 10^{12}$ G.   

The collapse of the initial iron core is simulated using the realistic 
nuclear Shen EOS \cite{shen98}, 
and the electron fraction (Ye) and entropy evolution following the parametrization of
neutrino physics by \citet{lie05}.  We neglect neutrino transport, and 
we plan such improvement to future work.  

The wavelength of the unstable mode is
directly proportional to the magnetic field strength.  Taking a conservative case,
for the typical pulsar field strength $\sim 10^{12}$ G, the most unstable
wavelength is $\sim 0.18 km (\frac{B}{10^{12}[G]}) (\frac{1000 [rad/s]}{\Omega}) 
(\frac{10^{13}}{\rho [g/cm^{3}]})^{0.5}$.  By giving 10 grid zones per wavelength to
resolve the MRI growth, $\sim 20$ m resolution is required to cover 
between $\sim 10 - $ few $\times 100$ km in the post--collapse environment.
It is unrealistic to simulate the iron core 
collapse with such a fine spatial resolution.  Therefore, we
first simulate the collapse of the iron core with n$_r$ = 256 and n$_{\theta}$ = 64, 
and 0 < radius < 6,800 km and 0 < $\theta$ < $\frac{\pi}{2}$ (base simulation).
The smallest radial resolution in the base simulation was $\sim 300$ m.
At 30 msec after the core bounce in the base simulation, we remap the profiles 
to a new mesh with n$_r$ = 1024 or 2048 and n$_{\theta}$ = 1024 or 2048 
(std or stdx2 run), and 
12 < radius < 68 km and $\frac{\pi}{6}$ < $\theta$ < $\frac{5\pi}{6}$.  
The smallest spatial resolution was
11 and 5.5 m respectively for std and stdx2 runs.

\section{Results}

%%%%%%%%%%%%%%%%%%%%%%%%%%%%%%%%%%%%%%%%%%%%
%% Sample figure:
%%
%% The option [height=...] scales the picture to the given height,
%% without it it would be printed at its nominal size
%%%%%%%%%%%%%%%%%%%%%%%%%%%%%%%%%%%%%%%%%%%%

\begin{figure}[ht]
\begin{minipage}[b]{0.5\linewidth}
\centering
\includegraphics[scale=0.35]{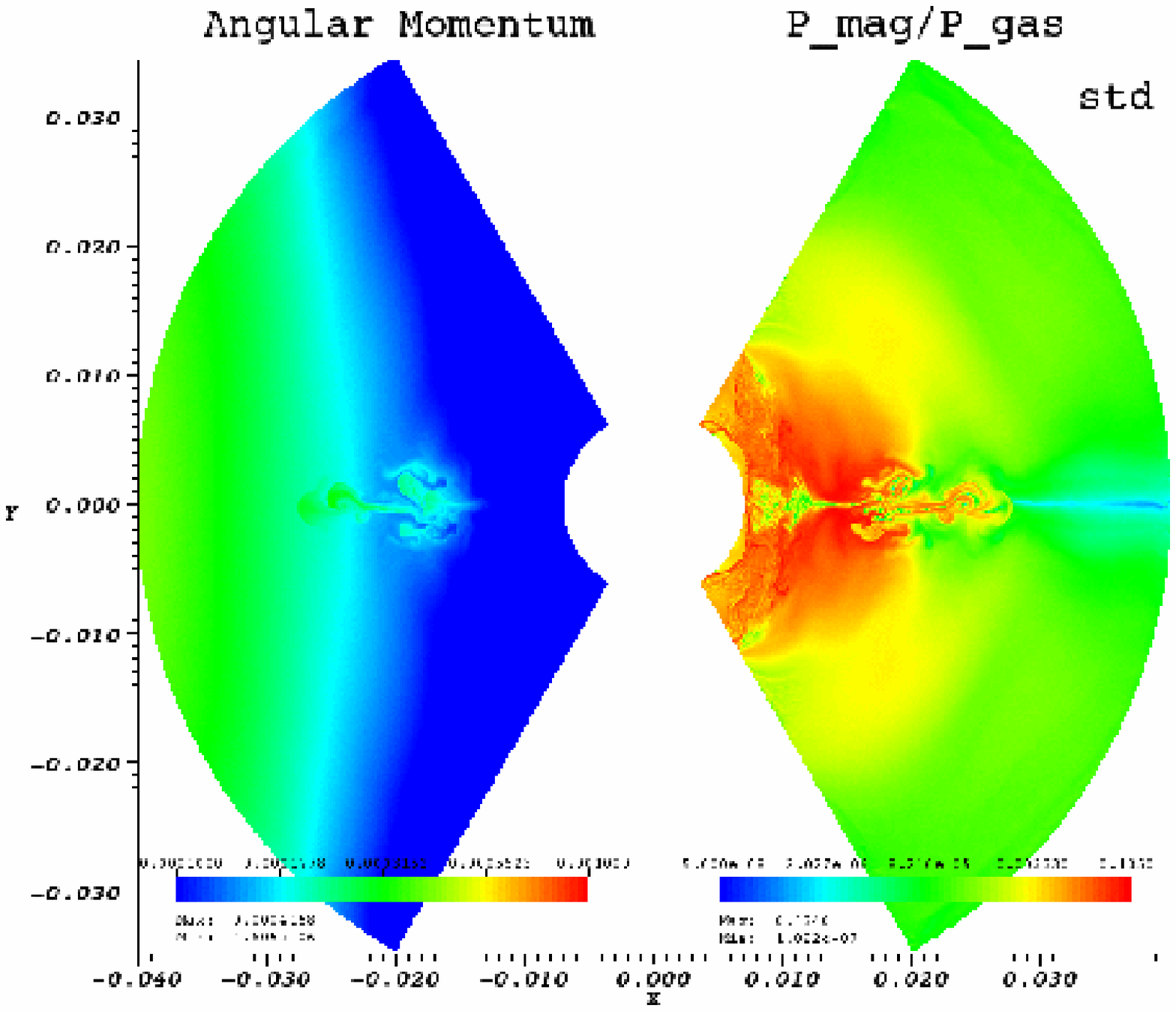}
\end{minipage}
\hspace{0.5cm}
\begin{minipage}[b]{0.5\linewidth}
\centering
\includegraphics[scale=0.35]{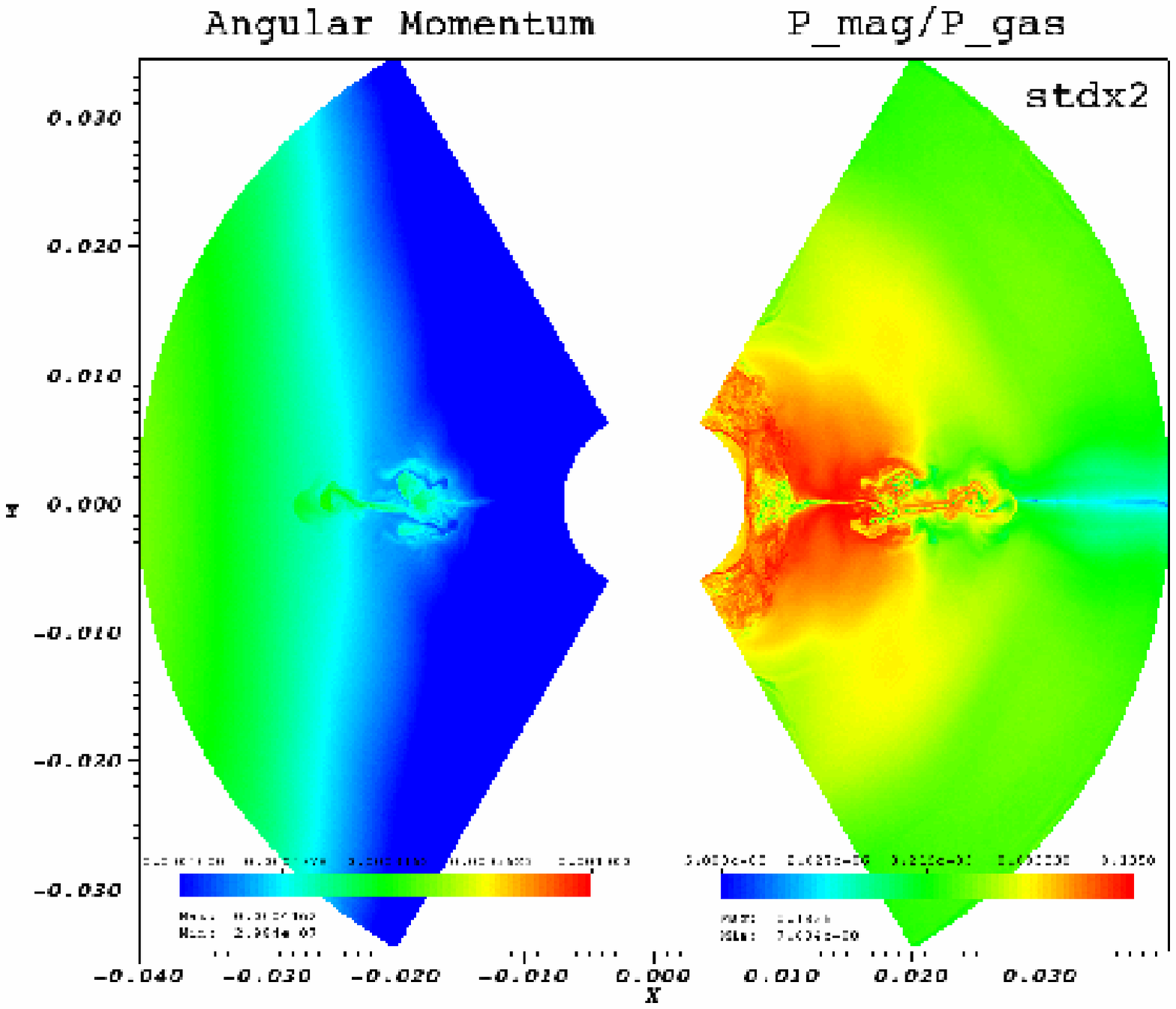}
\caption{Angular momentum and ratio between magnetic pressure and
gas pressure for std (left panel) and stdx2 (right panel) runs at
9 msec after remapping.  The simulated region is $12 - 68$ km in radius
and $0 - \frac{\pi}{2}$ in $\theta$.}
\label{fig:figure1}
\end{minipage}
\end{figure}

\begin{figure}[ht]
\begin{minipage}[b]{0.3\linewidth}
\centering
\includegraphics[scale=0.25]{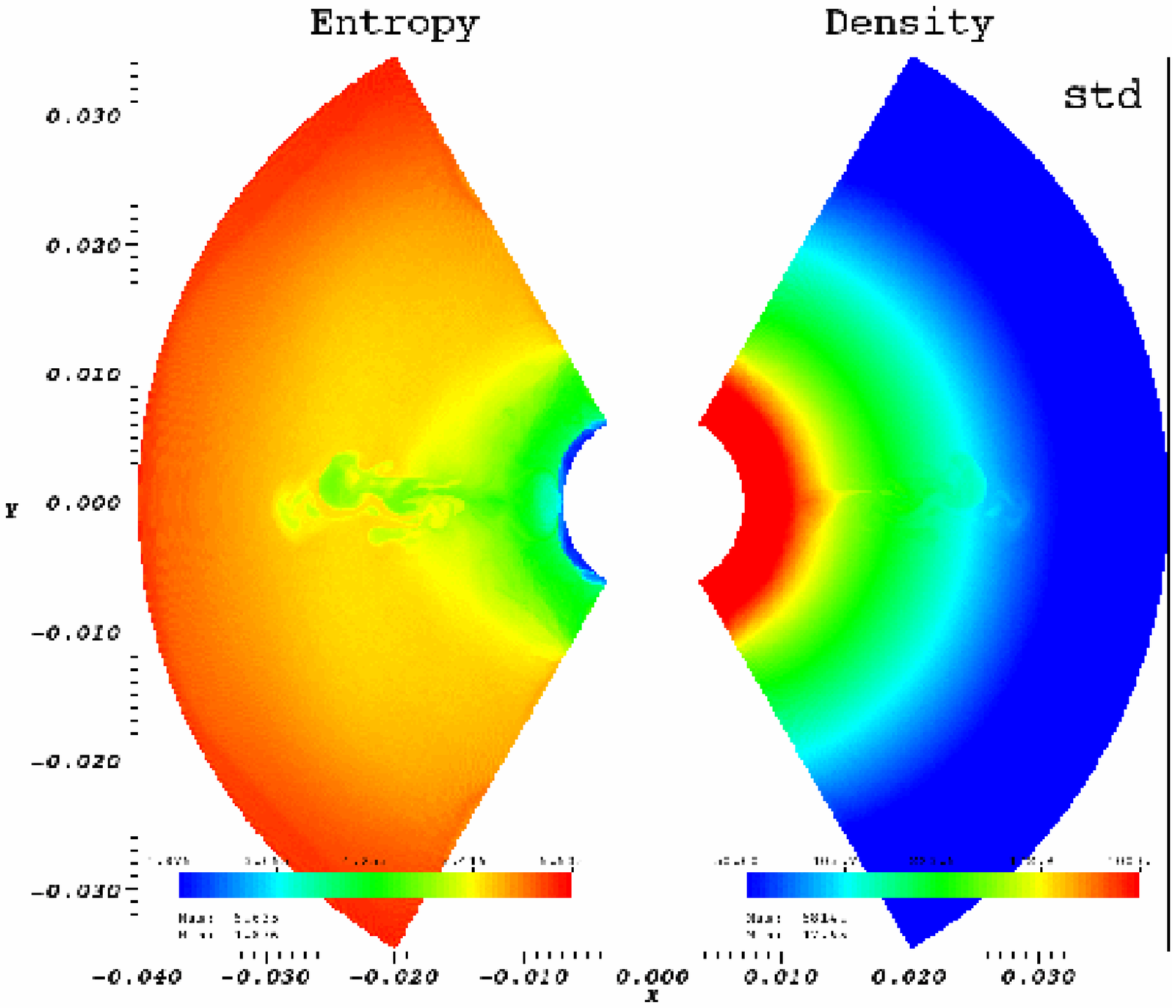}
\end{minipage}
\hspace{0.5cm}
\begin{minipage}[b]{0.3\linewidth}
\centering
\includegraphics[scale=0.25]{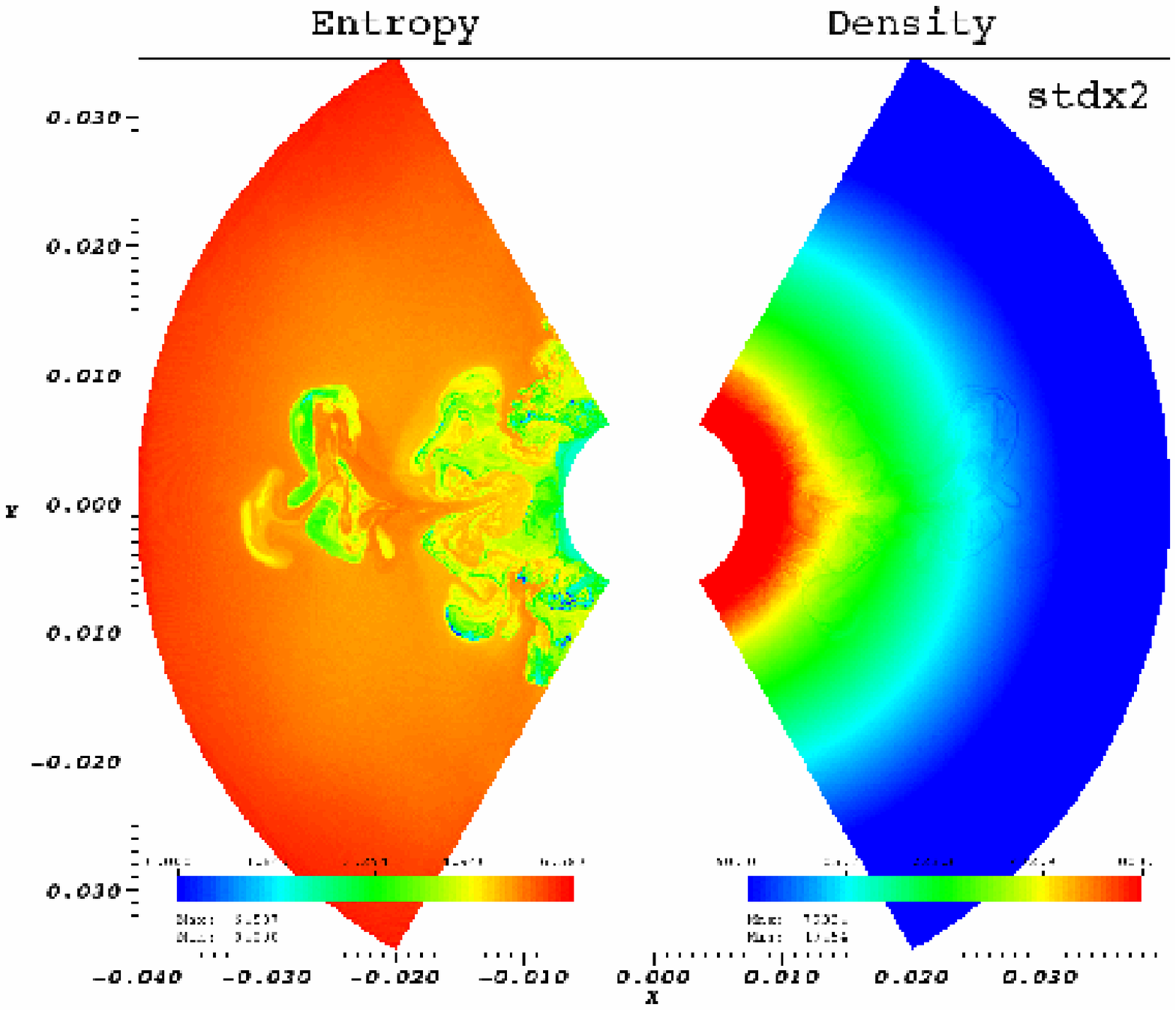}
\end{minipage}
\hspace{0.5cm}
\begin{minipage}[b]{0.3\linewidth}
\centering
\includegraphics[scale=0.25]{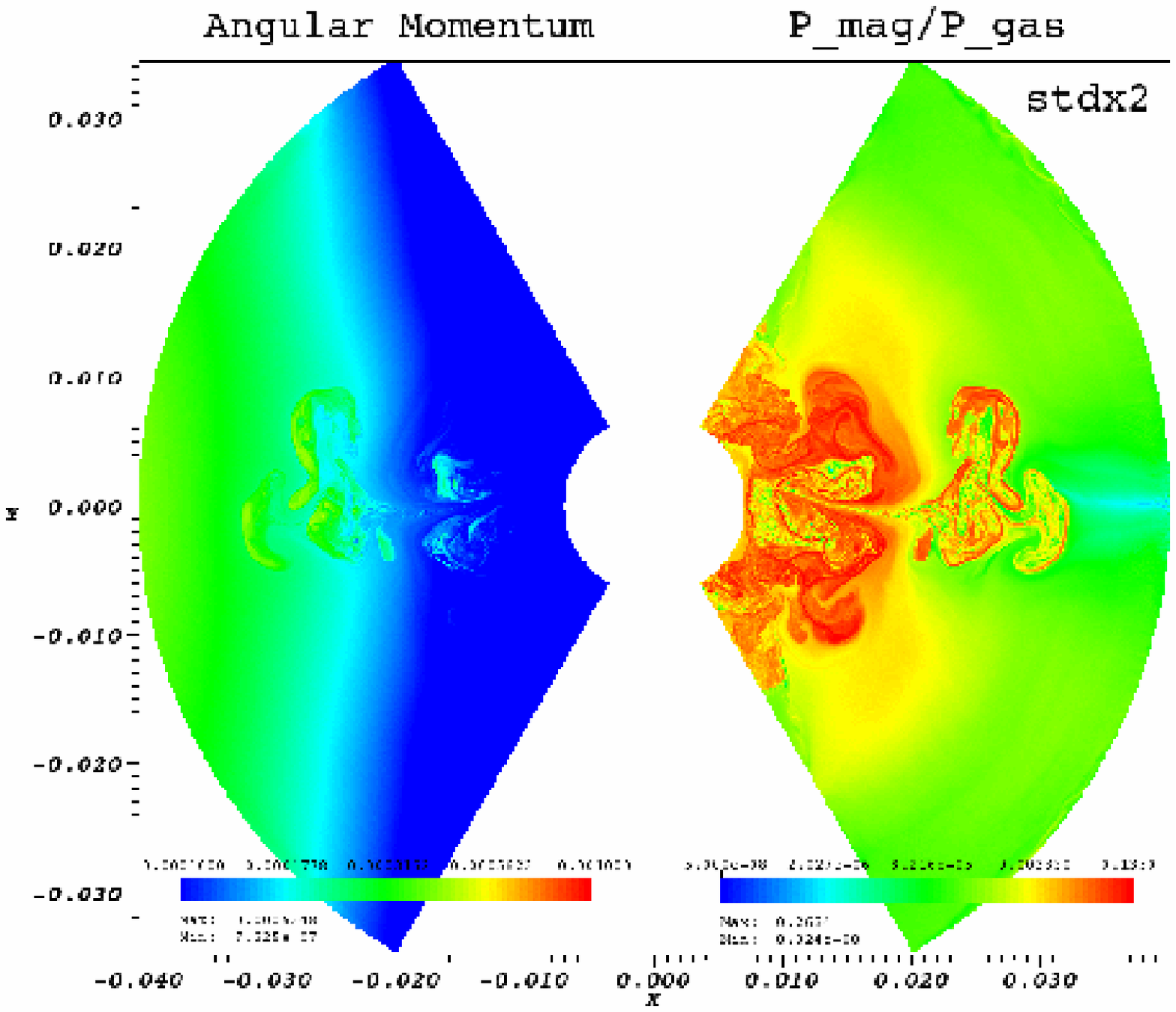}
\caption{Entropy and mass density for std (left) and stdx2 (middle) 
runs at 15 msec after remapping.  The right panel is the same with
Figure \ref{fig:figure1} for stdx2 run at 15 msec after remapping.}
\label{fig:figure2}
\end{minipage}
\end{figure}

%\begin{figure}[ht]
%\begin{minipage}[b]{0.5\linewidth}
%\centering
%\includegraphics[scale=1.2]{figures/Figure6.eps}
%\end{minipage}
%\hspace{0.5cm}
%\begin{minipage}[b]{0.5\linewidth}
%\centering
%\includegraphics[scale=1.2]{figures/Figure3.eps}
%\caption{Entropy and mass density for std (left panel) and stdx2 (right panel) runs 
%at 15 msec after remapping.}
%\label{fig:figure3}
%\end{minipage}
%\end{figure}

After the iron core collapsed, its initial magnetic fields were amplified up to
$\sim 10^{15}$ G, and the core was spun up to $\sim 2000$ [rad/s].
At the time of remapping,  
the most unstable wavelength of the MRI were few km.  The resolution 
obtained by both std and stdx2 runs are sufficient to resolve
the MRI growth.  Indeed, during the first 9 msec after the remapping, the evolution
of std and stdx2 runs are very similar (Figure \ref{fig:figure1}).  In both runs, the MRI unstable
region appears near the equatorial region, and magnetic bubbles rises in
the cylindrical radius direction, instead of spherical radius direction.  
At 15 msec after the remapping, the two runs shows differences (Figure \ref{fig:figure2}); higher
resolution run (stdx2) display more vigorous overturns and stronger outward transport
of entropy and angular momentum.  
The MRI unstable region kept filling up the simulated area until the end 
of the std run (45 msec).

Our results confirm that the MRI is unstable in the core collapse environment.
The qualitative results presented here are also
in agreement with \citet{aki03} and \citet{ober09}.  It is noted that the additional mixing of entropy 
caused by magnetic instability may affect the neutrino emission opacity.  This point
should be further investigated with future simulations with neutrino
transport.

%%%%%%%%%%%%%%%%%%%%%%%%%%%%%%%%%%%%%%%%%%%%%%%%
%% BACKMATTER
%%%%%%%%%%%%%%%%%%%%%%%%%%%%%%%%%%%%%%%%%%%%%%%%

\begin{theacknowledgments}
S.~A. is supported by grants from the Kavli Institute for Particle Astrophysics
and Cosmology (KIPAC) through a SciDAC Postdoctoral Fellowship.  S.~A. and J.~S. thank Christian 
D. Ott for providing the Shen EOS table.
\end{theacknowledgments}

%%%%%%%%%%%%%%%%%%%%%%%%%%%%%%%%%%%%%%%%%%%%%%%%
%% The bibliography can be prepared using the BibTeX program or
%% manually.
%%
%% The code below assumes that BibTeX is used.  If the bibliography is
%% produced without BibTeX comment out the following lines and see the
%% aipguide.pdf for further information.
%%
%% For your convenience a manually coded example is appended
%% after the \end{document}
%%%%%%%%%%%%%%%%%%%%%%%%%%%%%%%%%%%%%%%%%%%%%%%%

%%%%%%%%%%%%%%%%%%%%%%%%%%%%%%%%%%%%%%%%%%%%%%%%
%% You may have to change the BibTeX style below, depending on your
%% setup or preferences.
%%
%%
%% For The AIP proceedings layouts use either
%%%%%%%%%%%%%%%%%%%%%%%%%%%%%%%%%%%%%%%%%%%%

\bibliographystyle{aipproc}   % if natbib is available
%\bibliographystyle{aipprocl} % if natbib is missing

%%%%%%%%%%%%%%%%%%%%%%%%%%%%%%%%%%%%%%%%%%%
%% You probably want to use your own bibtex database here
%%%%%%%%%%%%%%%%%%%%%%%%%%%%%%%%%%%%%%%%%%%
\bibliography{ms}

%%%%%%%%%%%%%%%%%%%%%%%%%%%%%%%%%%%%%%%%%%%
%% Just a reminder that you may have to run bibtex
%% All of it up to \end{document} can be removed
%% if you don't like the warning.
%%%%%%%%%%%%%%%%%%%%%%%%%%%%%%%%%%%%%%%%%%%
\IfFileExists{\jobname.bbl}{}
 {\typeout{}
  \typeout{******************************************}
  \typeout{** Please run "bibtex \jobname" to optain}
  \typeout{** the bibliography and then re-run LaTeX}
  \typeout{** twice to fix the references!}
  \typeout{******************************************}
  \typeout{}
 }

\end{document}